\documentclass[%
reprint,
amsmath,amssymb,
aps,
prb,
]{revtex4-1}
\usepackage[dvips]{color}
\usepackage{amsfonts}
\usepackage{xcolor}
\usepackage{amsmath}
\usepackage{amssymb}
\usepackage{graphicx}
\usepackage{mathrsfs}
\usepackage{bbm}
\usepackage{bm}
\usepackage{multirow}
\usepackage[sort&compress]{natbib}
\usepackage{dcolumn}%

\begin{document}

\newcommand{\dd}{d}
\newcommand{\pd}{\partial}
\newcommand{\myU}{\mathcal{U}}
\newcommand{\myr}{q}
\newcommand{\Urho}{U_{\rho}}
\newcommand{\myalpha}{\alpha_*}
\newcommand{\bd}[1]{\mathbf{#1}}
\newcommand{\Eq}[1]{Eq.~(\ref{#1})}
\newcommand{\Eqn}[1]{Eq.~(\ref{#1})}
\newcommand{\Eqns}[1]{Eqns.~(\ref{#1})}
\newcommand{\Eqnsa}[1]{Eqns.~(\ref{#1}}
\newcommand{\Figref}[1]{Fig.~\ref{#1}}
\newtheorem{theorem}{Theorem}
\newcommand{\me}{\textrm{m}_{\textrm{e}}}
\newcommand{\sgn}{\textrm{sign}}
\newcommand*{\bfrac}[2]{\genfrac{\lbrace}{\rbrace}{0pt}{}{#1}{#2}}
\newcommand{\figsize}{8.6cm}
\newcommand{\I}{\textrm{I}}
\newcommand{\II}{\textrm{II}}
\newcommand{\III}{\textrm{III}}
\newcommand{\IV}{\textrm{IV}}
\newcommand{\V}{\textrm{V}}
\newcommand{\VI}{\textrm{VI}}
\newcommand{\VII}{\textrm{VII}}
\newcommand{\VIII}{\textrm{VIII}}

\DeclareBoldMathCommand\boldl{\left(} 
\DeclareBoldMathCommand\boldr{\right)}

\def\sigh#1{%
\pmb{\left( \vphantom{#1}\right.}%
#1%
\pmb{\left. \vphantom{#1}\right)}}

\preprint{To be submitted to PRX}

\title{Band warping, band non-parabolicity and Dirac points in fundamental lattice and electronic structures}

\author{Lorenzo Resca}%
 \email{resca@cua.edu}
 
\author{Nicholas A. Mecholsky}
 \thanks{Corresponding Author}
 \email{nmech@vsl.cua.edu}
 
\author{Ian L. Pegg}
 \email{ianp@vsl.cua.edu}
\affiliation{%
 Department of Physics and Vitreous State Laboratory\\
The Catholic University of America\\
Washington, DC 20064}

\date{\today}

\begin{abstract}
We demonstrate from a fundamental perspective the physical and mathematical origins of band warping and band non-parabolicity in electronic and vibrational structures. Remarkably, we find a robust presence and connection with pairs of topologically induced Dirac points in a primitive-rectangular lattice using a $p$-type tight-binding approximation. We provide a transparent analysis of two-dimensional primitive-rectangular and square Bravais lattices whose basic implications generalize to more complex structures. Band warping typically arises at the onset of a singular transition to a crystal lattice with a larger symmetry group, suddenly allowing the possibility of irreducible representations of higher dimensions at special symmetry points in reciprocal space. Band non-parabolicities are altogether different higher-order features, although they may merge into band warping at the onset of a larger symmetry group. Quite separately, although still maintaining a clear connection with that merging, band non-parabolicities may produce pairs of conical intersections at relatively low-symmetry points. Apparently, such conical intersections are robustly maintained by global topology requirements, rather than any local symmetry protection. For two $p$-type tight-binding bands, we find such pairs of conical intersections drifting along the edges of restricted Brillouin zones of primitive-rectangular Bravais lattices as lattice constants vary relatively, until they merge into degenerate warped bands at high-symmetry points at the onset of a square lattice. The conical intersections that we found appear to have similar topological characteristics as Dirac points extensively studied in graphene and other topological insulators, although our conical intersections have none of the symmetry complexity and protection afforded by the latter more complex structures.
\end{abstract}

\maketitle

\section{\label{sec:level1}Introduction}
In this paper we analyze the physical and mathematical origin of band warping using two fundamental and complementary approaches to calculate electronic band structures, namely, the tight-binding (TB) method and the nearly-free-electron (NFE) model. More generally, we demonstrate that for the majority of non-degenerate bands one should not expect band warping at energy extrema, irrespective of whether band non-parabolicity may or may not be prominent. Band non-parabolicity should not be confused with band warping. Band non-parabolicity derives from terms of order higher than quadratic in a multi-dimensional Taylor series expansion. By contrast, in the case of band warping, there is no possibility of performing a multi-dimensional Taylor series approximation at the quadratic order already. Now, band non-parabolicity becomes more and more evident when greater and greater coefficients of higher-order terms develop in its Taylor series expansion, possibly as a result of stronger and stronger interactions with other bands approaching in energy the non-parabolic band. Nevertheless, insofar as a non-parabolic band remains non-degenerate at an energy extremum, it typically remains non-warped therein. However, if a singular transition from a lower symmetry lattice to a higher symmetry lattice occurs, possibly as a result of variations in lattice constant parameters, a highly singular transformation may correspondingly ensue. Namely, highly non-parabolic bands may forcibly become degenerate with one another at an isolated point in reciprocal space as a result of greater symmetry requirements. Then, though not invariably, band non-parabolicity coefficients may transform singularly into a warped band structure at the isolated point of degeneracy. Smoothness of the touching bands is destroyed at the quadratic order of curvature already. Ultimately, the warped singularity is a consequence of the singular transition to a larger symmetry group, allowing the possibility of irreducible representations of higher dimensions at special symmetry points in reciprocal space.
\begin{figure}[!hb]
	\begin{center}
 	\includegraphics[width=\figsize]{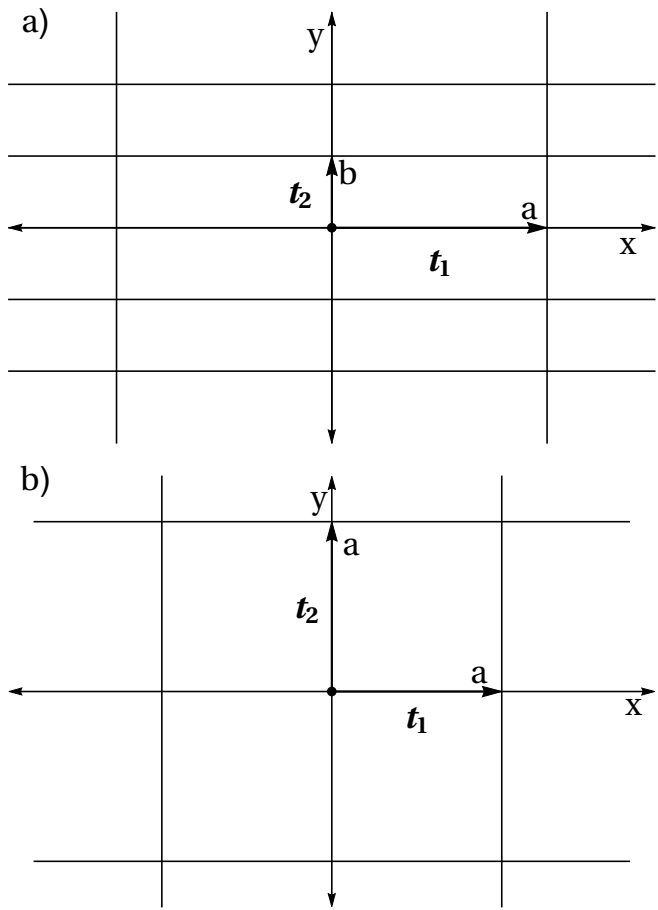}\\
      \caption{\label{fig:fig1} Primitive-rectangular (a) and square (b) Bravais lattices.}
\end{center}
\end{figure}

Characteristics of ``warping'' are not exclusive to electronic or phononic band structures in crystals. Perhaps as a deceivingly simple example, let us consider a potential function in classical mechanics or electrostatics of the form 
\begin{equation}\label{eqn:example}
z = f(x,y) = \frac{x^4 + y^4}{x^2 + y^2},
\end{equation}
having further defined $z=0$ at the origin. This function could represent a frictionless surface onto which a point-particle slides under a uniform gravitational field in the negative $z$-direction. One can easily verify that $f(x,y)$ represents a perfectly valid single-valued potential everywhere, yielding a conservative and irrotational force field, $\bd{F} = - \nabla{f}$, everywhere. In fact, $f(x,y)$ is continuous with continuous partial derivatives everywhere. It has a single absolute minimum at the origin, with a zero first-differential, representing a tangent horizontal plane at the origin. However, $f(x,y)$ has discontinuous partial second-order derivatives at the origin, exclusively. Thus, $f(x,y)$ does not admit a second-differential at the origin, exclusively. This means that no smooth curvature can be defined for $f(x,y)$ at the origin, exclusively. Correspondingly, no Taylor series expansion can be performed for $f(x,y)$ at the origin, beyond the trivially zero first-order term. Consequently, no theory of small oscillations is at all possible for such a simple potential, and countless more of that sort: a fact rarely mentioned in standard textbooks. Correspondingly, complex chaotic and non-chaotic orbits develop at all scales around the minimum. Functions of this sort, readily extended to three dimensions, may well represent interactions between atoms in molecules or crystals, excluding even the possibility of harmonic vibrations and quantization of phonons. 

Why then should any such seemingly reasonable function be excluded from physical consideration? Let us consider more suitable systems of curvilinear coordinates, such as cylindrical $(z,r,\theta)$ coordinates, in which our token example is expressed as
\begin{equation}\label{eqn:example cylindrical coordinates}
z = f(x,y) = r^2 \left( \frac{3}{4} + \frac{1}{4} \cos ( 4 \theta) \right) .
\end{equation}
This expression shows that our function is exactly quadratic in $r$, the radial coordinate distance from the $z$-axis, but it maintains a periodically undulating angular term in $(4 \theta)$ for any non-zero distance, no matter how small. 

It turns out that in a quantum mechanical theory of crystals, though for reasons that are not immediately obvious, energy functions of the kind that we have just described, lacking smooth curvature at isolated critical points in reciprocal space, typically occur under certain conditions. This solid state feature, traditionally called ``band warping" of electronic or phononic energy structures, was discovered and it has been investigated for quite some time. Original work began in the 50's.\cite{LuttingerKohn1955Motion, Dresselhaus, laxmavroides, Phillips56, Kane1956, Kane1957band} Only recently, however, some authors have provided a more mathematically precise and applicable description of band warping based on the idea and formalism of an angular effective mass, which has lead to better treatments of transport, optical, and other fundamental properties of materials.\cite{MRPF, MRPFII, janssen2016precise, gonze2016ABINITwarping, parker2015benefits} 

\begin{figure}[!hb]
	\begin{center}
 	\includegraphics[width=\figsize]{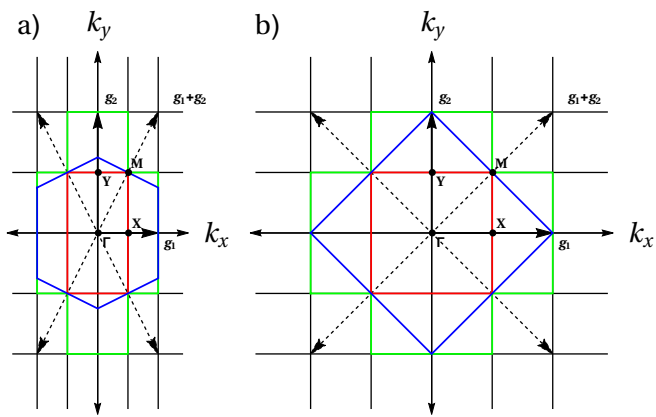}\\
      \caption{\label{fig:fig2} Primitive-rectangular (a) and square (b) reciprocal lattices and their first three Brillouin zones, colored in red, blue, and green, respectively.}
\end{center}
\end{figure}

 In order to demonstrate unequivocally the origin of this basic phenomenon, let us begin by considering two fundamental crystal structures, namely, those of two-dimensional primitive-rectangular and square Bravais lattices, shown in \Figref{fig:fig1} a) and b). Their two primitive lattice vectors are $\bd{t}_1 = a (1,0)$ and $\bd{t}_2 = b (0,1)$ for the primitive-rectangular lattice, while $a = b$ become identical for the square lattice. A generic lattice point is represented by $\boldsymbol{\tau} = n_1 \bd{t}_1 + n_2 \bd{t}_2$, where $n_1$ and $n_2$ represent any two integers. Reciprocal lattices and vectors are shown in \Figref{fig:fig2} a) and b). Their two primitive reciprocal lattice vectors are $\bd{g}_1 = \tfrac{2 \pi}{a} (1,0)$ and $\bd{g}_2 = \tfrac{2 \pi}{b} (0,1)$.  For the square lattice, $a = b$ become again identical. A generic reciprocal-lattice point is represented by the vector $\bd{G} = m_1 \bd{g}_1 + m_2 \bd{g}_2$, where $m_1$ and $m_2$ represent any two integers. Brillouin zones (BZ) are obtained by bisecting with straight lines all $\bd{G}$ vectors and by considering the resulting nested minimal polygonal enclosures, each comprising a unit of the basic area $|\bd{g}_1 \times \bd{g}_2 | = \tfrac{(2 \pi)^2}{a b}$. The first, second, and third BZ are highlighted in \Figref{fig:fig2} with red, blue, and green colors.\cite{Bassani, GP, Ashcroft} We shall proceed to calculate fundamental characteristics of associated band structures. We will discover a surprising richness of possibilities at the most fundamental level, including pair formation of \textit{conical intersections}, heretofore considered as Dirac points peculiar to graphene and other complex topological insulators.\cite{CastroNoselovGeimRMP2009, BansilLinDasRMP2016}
\section{\label{sec:tightbinding}Tight-binding method for $p$-type localized orbitals in a primitive-rectangular lattice}
Tight-binding (TB) approximations provide perhaps the most basic method of calculating electronic energy band structures. TB is most applicable to bands originating from core or valence electrons tightly bound to their atoms, typically forming insulating or semiconducting materials, such as molecular solids, covalently bonded compounds, or ionic crystals. The prototypical example that we consider here consists of two $p$-type valence bands, retaining a relatively small number of empirical TB parameters. Let us assume the existence of an orthonormal set of Wannier functions $p_x (\bd{r} - \boldsymbol{\tau})$ and $p_y (\bd{r} - \boldsymbol{\tau})$ in a primitive-rectangular Bravais lattice. Let us then form two corresponding Bloch sums and expand a crystal one-electron wave function $\Psi(\bd{k},\bd{r})$ on that basis. A standard procedure\cite{Bassani, GP} leads to a 2x2 determinant equation containing the matrix elements
\begin{subequations}\label{eqn:TB}
\begin{align}
M_{xx} (\bd{k}) &= V^{(\I)} (pp\pi) \cdot 2 \cos (k_y b) + V^{(\II)} (pp\sigma) \cdot 2 \cos (k_x a) \nonumber\\ &+ \left( V^{(\III)} (pp\sigma) \frac{a^2}{a^2 + b^2} + V^{(\III)} (pp\pi) \frac{b^2}{a^2 + b^2} \right) \nonumber\\ &\times 4 \cos (k_x a) \cos (k_y b) \\
M_{yy} (\bd{k}) &= V^{(\I)} (pp\sigma) \cdot 2 \cos (k_y b) + V^{(\II)} (pp\pi) \cdot 2 \cos (k_x a) \nonumber\\ &+ \left( V^{(\III)} (pp\sigma) \frac{b^2}{a^2 + b^2} +  V^{(\III)} (pp\pi) \frac{a^2}{a^2 + b^2} \right) \nonumber\\ &\times 4 \cos (k_x a) \cos (k_y b) \\
M_{xy} (\bd{k}) &= M_{yx} (\bd{k}) = \left( V^{(\III)} (pp\pi) -  V^{(\III)} (pp\sigma) \right) \nonumber\\ &\times \frac{4 a b}{a^2 + b^2} \sin (k_x a) \sin (k_y b).
\end{align}
\end{subequations}

In \Eqnsa{eqn:TB}a-c), the superscripts $(\I)$, $(\II)$, and $(\III)$ refer to first-, second-, and third-nearest neighbors, located at $\pm \bd{t}_2$, $\pm \bd{t}_1$, $\pm \bd{t}_1 \pm \bd{t}_2$ and $\pm \bd{t}_1 \mp \bd{t}_2$, respectively. In \Eqnsa{eqn:TB}a-c), $V(pp\sigma)$, typically positive for attractive potentials, and $V(pp\pi)$, typically negative and smaller for attractive potentials, refer to two-center interaction integral parameters originally introduced by Slater and Koster (see Ref.\ \onlinecite{GP}, p.\ 146-150, for example). Further TB parameters can and will be considered later, but they are inessential for the prototypical formulation that we are going to consider first.

\begin{figure}[!hb]
	\begin{center}
 	\includegraphics[width=\figsize]{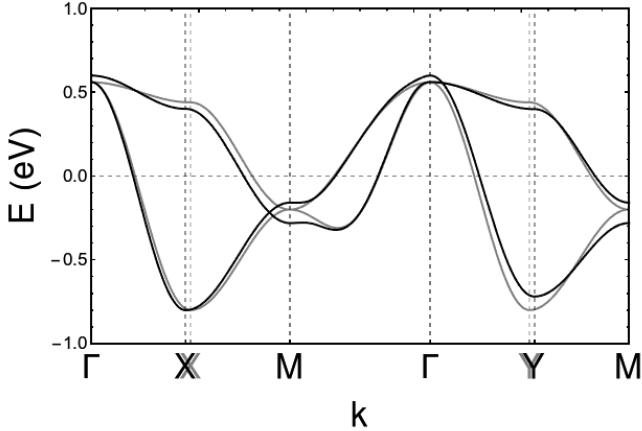}\\
      \caption{\label{fig:fig3} Bands along a path around and across the first BZ of primitive-rectangular and square reciprocal lattices derived from a tight-binding model with two $p$-type localized basis functions. Nominal values for the corresponding interaction parameters are $V^{(\I)} (pp\sigma)=0.25$, $V^{(\II)} (pp\sigma)=0.24$, $V^{(\III)} (pp\sigma)=0.1$, and $V^{(\I)} (pp\pi)=-0.06$, $V^{(\II)} (pp\pi)=-0.03$, $V^{(\III)} (pp\pi)=-0.01$ for the primitive-rectangular case (black), and $V^{(\I)} (pp\sigma)=0.25$, $V^{(\II)} (pp\sigma)=0.25$, $V^{(\III)} (pp\sigma)=0.1$, and $V^{(\I)} (pp\pi)=-0.06$, $V^{(\II)} (pp\pi)=-0.06$, $V^{(\III)} (pp\pi)=-0.01$ for the square case (gray). The lattice constants have nominal values of $a=1.1$ and $b=1$ for the primitive-rectangular case and $a=b=1$ for the square case.}
%
\end{center}
\end{figure}

With this basic $p$-type TB model, we thus arrive at the following two, typically non-degenerate, bands at each $\bd{k}$-point in the first BZ of the primitive-rectangular reciprocal lattice:
\begin{align}\label{eqn:detsol}
E_{\pm} (\bd{k}) &= \frac{1}{2} \left( M_{xx} (\bd{k}) + M_{yy} (\bd{k}) \phantom{\frac{1}{2}} \right. \nonumber\\
&\phantom{=} \pm \left. \sqrt[]{\left( M_{xx} (\bd{k}) - M_{yy} (\bd{k})\right)^2 + 4 M^2_{xy} (\bd{k})} \right) .
\end{align}
The energy difference or gap, $E_g$, that occurs at $\bd{k}=0$, or $\Gamma$-point, generates a positive constant, $E_g^2 = \left(M_{xx} (\bd{0}) - M_{yy} (\bd{0}) \right)^2 > 0$, under the square root. This positive constant allows to perform a Taylor series expansion of the square root to all orders in $\bd{k}$, including a quadratic expansion in $\bd{k}$ around the relative minimum in energy difference at $\Gamma$. No band warping can thus arise. For nominal values of the interaction parameters (see Ref.\ \onlinecite{GP}, p.\ 152, for example) such as $V^{(\I)} (pp\sigma)=0.25$, $V^{(\II)} (pp\sigma)=0.24$, $V^{(\III)} (pp\sigma)=0.1$, and $V^{(\I)} (pp\pi)=-0.06$, $V^{(\II)} (pp\pi)=-0.03$, $V^{(\III)} (pp\pi)=-0.01$, the electronic band structure along the path $\Gamma \rightarrow X \rightarrow M \rightarrow \Gamma \rightarrow Y \rightarrow M$, across and around the edges of the restricted BZ in reciprocal space, is shown in black in \Figref{fig:fig3}. Notice, remarkably, that there are band degeneracies at isolated points in the BZ. Any such degeneracy must be regarded as ``accidental," because the point-group of symmetry, $C_{2 v}$, of a primitive-rectangular lattice provides no irreducible representation of dimension greater than one.\cite{Bassani, landauLifshitzQM} These ``accidental'' degeneracies will later be shown to represent topologically required \textit{conical intersections} of utmost theoretical interest.

One may wonder whether in general \textit{non-degenerate} energy bands may be warped at isolated points of extremal energy, i.e., minima, maxima, or saddle points in the BZ, for any crystal structure or model. We have investigated that question to considerable extent. We found that that is possible, but only under special conditions. For example, we can refine the TB model that we have just described for a primitive-rectangular lattice to include $S(pp\sigma)$ and $S(pp\pi)$ overlap integrals among $p$-type atomic-like orbitals, rather than presuming orthonormal Wannier functions from the beginning. In that situation, after some laborious and not necessarily physical ``fine-tuning" of interaction and overlap parameters, non-degenerate energy bands can be fabricated as having at $\bd{k}=0$ an isolated critical point with non-smooth curvature, i.e., \textit{non-degenerate band warping}. Thus, warped non-degenerate electronic bands, similar in kind to the classical potential mentioned in the Introduction, are not impossible to realize with special materials, but any such occurrence should be regarded as rather fortuitous and hardly robust or generic. 

\section{\label{sec:tightbindingsquare}Tight-binding method for $p$-type localized orbitals in a square lattice}
The two-band $p$-type structure of a square lattice cannot be continuously derived from that of the primitive-rectangular lattice by letting $a$ and $b$ approach each other continuously. For the square lattice, the interaction parameters $V^{(\I)} (pp\sigma) = V^{(\II)} (pp\sigma)$ and $V^{(\I)} (pp\pi) = V^{(\II)} (pp\pi)$ must become respectively identical because of the square symmetry, thus profoundly transforming the two-band TB matrix elements as
\begin{subequations}\label{eqn:TB2}
\begin{align}
M_{xx} (\bd{k}) &= V^{(1)} (pp\sigma) \cdot 2 \cos (k_x a) + V^{(1)} (pp\pi) \cdot 2 \cos (k_y a) \nonumber\\ &+ \left( V^{(2)} (pp\sigma) + V^{(2)} (pp\pi) \right) \nonumber\\ &\times 2 \cos (k_x a) \cos (k_y a) \\
M_{yy} (\bd{k}) &= V^{(1)} (pp\pi) \cdot 2 \cos (k_x a) + V^{(1)} (pp\sigma) \cdot 2 \cos (k_y a) \nonumber\\ &+ \left( V^{(2)} (pp\sigma) +  V^{(2)} (pp\pi) \right) \nonumber\\ &\times 2 \cos (k_x a) \cos (k_y a) \\
M_{xy} (\bd{k}) &= M_{yx} (\bd{k}) = \left( V^{(2)} (pp\pi) -  V^{(2)} (pp\sigma) \right) \nonumber\\ &\times 2 \sin (k_x a) \sin (k_y a).
\end{align}
\end{subequations}

In \Eqnsa{eqn:TB2}a-c) the superscripts (1) and (2) now refer to the first- and second-nearest neighbors located at $\pm \bd{t}_{1}$ and $\pm \bd{t}_{2}$, and $\pm \bd{t}_{1} \pm \bd{t}_{2}$ and $\pm \bd{t}_{1} \mp \bd{t}_{2}$, respectively, as a pair of quadruplets. 

The electronic band structure for a square lattice along the path $\Gamma \rightarrow X \rightarrow M \rightarrow \Gamma \rightarrow Y \rightarrow M$ in reciprocal space is shown in gray in \Figref{fig:fig3}. Formally, \Eq{eqn:detsol} still represents the determinant solution for the two bands, $E_{\pm} (\bd{k})$, for the square lattice. However, the energy gap, $E_g$, vanishes identically at $\Gamma$ and $M$ points for the square lattice. In fact, at $\bd{k}=0$, we have
\begin{equation}
E_g^2 = \left( M_{xx}(\bd{0}) - M_{yy} (\bd{0}) \right)^2 \equiv 0 ,
\end{equation}
and also
\begin{equation}
M_{xy}^2 (\bd{0}) \equiv 0 .
\end{equation}
Thus, the lowest term in the Taylor series expansion in $\bd{k}$ of the discriminant under the square root in \Eq{eqn:detsol} is only of 4th-order, hence,
\begin{align}\label{eqn:TB3}
&\sqrt[]{\left[ M_{xx} (\bd{k}) - M_{yy} (\bd{k}) \right]^2 + 4 M^2_{xy} (\bd{k}) } \nonumber\\ &\simeq \left\lbrace \left[V^{(1)} (pp\sigma) - V^{(1)} (pp\pi)\right]^2 \cdot a^4 \left( k_x^2 - k_y^2 \right)^2 \right. \nonumber\\  &+\left. 16 \left[ V^{(2)} (pp\sigma) - V^{(2)} (pp\pi) \right]^2 \cdot a^4 k_x^2 k_y^2 \right\rbrace^{1/2}.
\end{align}

We notice that there is no constant term under the square root in \Eq{eqn:detsol} when expanded as \Eq{eqn:TB3}. In fact, all terms are at least quartic in the Cartesian $\bd{k}$-components under the square root in \Eq{eqn:TB3}. Therefore, there cannot be any multi-dimensional Taylor series expansion of the square-root function around $\bd{k}=0$, unless we totally disregard second-nearest-neighbor interactions, i.e., we set especially $V^{(2)} (pp\sigma)=V^{(2)} (pp\pi) = 0$. In that case only the two bands merely contact one another smoothly at $\bd{k}=0$. Otherwise, and most generally, the two bands interact and distort one another, creating a point of contact deprived of any smooth curvature for either band, that is, an isolated point of \textit{degeneracy and band warping}. Notice that this degeneracy of the two $p$-like bands at $\bd{k}=0$ is required by the point-group of symmetry, $C_{4v}$, which has doubled in size for the square lattice, and it thus admits two-dimensional $p$-like irreducible representations.\cite{Bassani, landauLifshitzQM} 

We may also notice in \Figref{fig:fig3} that the two non-degenerate bands of the primitive-rectangular lattice that approach one another in the limit of a square lattice are represented by relative maxima in the proximity of the $\Gamma$-point. However, because of their vicinity and strong interaction, these two non-degenerate bands promptly develop considerable non-parabolicity. That contributes to their crossing with a \textit{conical intersection} at an isolated point near the $\Gamma$-point along the $\Gamma \rightarrow Y$ line: cf.\ \Figref{fig:fig3}. Later, that conical intersection will be more fully analyzed and displayed in \Figref{fig:fig5}. 
\begin{figure}[!hpt]
	\begin{center}
 	\includegraphics[width=8cm]{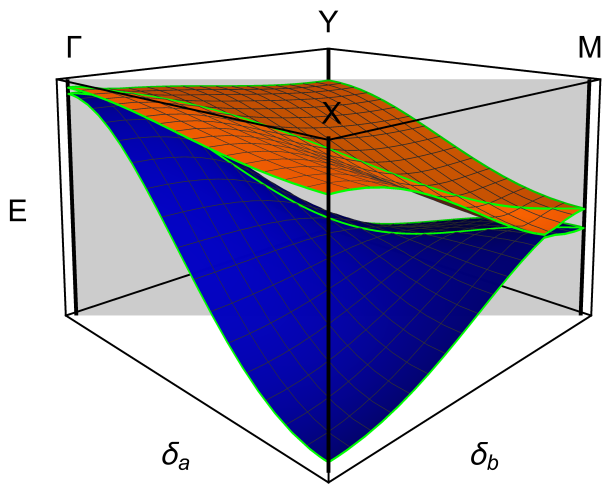}\\
    \caption{\label{fig:fig4} Two-dimensional representation of the two $p$-like TB bands in the restricted BZ of the primitive-rectangular lattice. A \textit{conical intersection} close to the $M$-point along the $M \rightarrow X$ line is clearly visible. Another \textit{conical intersection} close to $\Gamma$-point along the $\Gamma \rightarrow Y$ line lies mostly hidden behind the two bands in this figure.}
\end{center}
\end{figure}

A similar analysis can be performed at the $M$-point of the square lattice, where $\bd{k}=\tfrac{1}{2} \left( \bd{g}_1 + \bd{g}_2 \right) = \pi \left( \tfrac{1}{a}, \tfrac{1}{a} \right)$. In that case, the two non-degenerate bands of the primitive-rectangular lattice (approaching one another in the limit of a square lattice) are represented by saddle points at $M$. Again, those saddle-point bands promptly develop considerable non-parabolicity, leading to a second \textit{conical intersection} at an isolated point along the $M \rightarrow X$ line: cf.\ \Figref{fig:fig3}. This other conical intersection is fully displayed in the two-dimensional expanded view of the two $p$-like TB bands of the primitive-rectangular lattice shown in \Figref{fig:fig4}.

Remarkably, but understandably, as the primitive-rectangular lattice approaches the limit of a square lattice, this pair of \textit{conical intersections} approach, and ultimately merge with, the degenerate and warped band structures at $\Gamma$ and $M$ points, respectively. That merging mechanism is in fact intimately connected with the vanishing of any positive constant under a square root that is ultimately responsible for the creation of band warping at those critical points in reciprocal space.

We may confirm and quantify the presence of band warping by computing a band-warping parameter, $w$, introduced in Ref.\ \onlinecite{MRPF} to provide some measure of band warping for two- and three-dimensional energy surfaces at a critical point. Non-warped energy dispersions must have $w = 0$. By \textit{radially} expanding \Eqn{eqn:detsol} around $\Gamma$ or $M$ points to second order in the $r$-distance, we can explicitly compute the warping parameter for the primitive-rectangular and square cases sketched in \Figref{fig:fig3} and \Figref{fig:fig4}. In the primitive-rectangular case, the warping parameter is zero for both non-degenerate bands at $\Gamma$ and $M$ points, as expected. In the square case, however, the warping parameter is $w = -0.0498$ and $w = -0.308$ for the two degenerate bands at $\Gamma$.  At the $M$-point, the warping parameter is $w = -0.124$ and $w = 0.115$ at $\Gamma$ and $M$ points, respectively. Negative and positive signs in $w$ indicate generalized band maxima and minima, respectively. We may further recall that the warping parameter, $w$, provides some measure of how severely an energy surface is prevented from having smooth curvature or second-order differentiability at an isolated critical point.\cite{MRPF} In this instance, calculation of $w$ confirms the expected warping at $\Gamma$ and $M$ points for the two $p$-type TB bands of a square lattice that we have demonstrated analytically. For comparison, the band-warping parameter is $w = \tfrac{1}{3 \sqrt{2}} \approx 0.2357$ for the classical potential mentioned in the Introduction.

The prototypical situation demonstrated with this basic TB square-lattice model captures the essence of what has been shown in three-dimensional IV-IV, III-V, II-VI and I-VII cubic compounds for $p$-type bands of interest.\cite{Dresselhaus, laxmavroides, Kane1956, Kane1957band, Bassani, MRPF, MRPFII} We have further applied TB and $\bd{k} \cdot \bd{p}$ models to investigate the origin of band warping at degeneracy points in reciprocal space for those and other crystal structures and materials. We have generally confirmed that band warping is the most typical result of symmetry-induced band-degeneracy.

\section{\label{sec:conicalintersections}Conical intersections}
Our observation of a pair of conical intersections along the edges of the restricted BZ of the primitive-rectangular Bravais lattice for a basic $p$-type two-band TB model is truly remarkable and thus deserves at least a few more comments. In our case, the band degeneracy involved is ``accidental,'' rather than required by symmetry. Normally, two-dimensional surfaces cross each other along one-dimensional curves. On the other hand, one generally expects quantum-mechanical interactions to fully remove any ``accidental'' degeneracy or crossing.\cite{landauLifshitzQM} In our conical intersections, something happens in between. Apparently, the global topology of the lattice forces the two-dimensional bands to go underneath and re-emerge from one another at a pair of singular ``strangulated'' points. 

\begin{figure}[!hpt]
	\begin{center}
    \includegraphics[width=8cm]{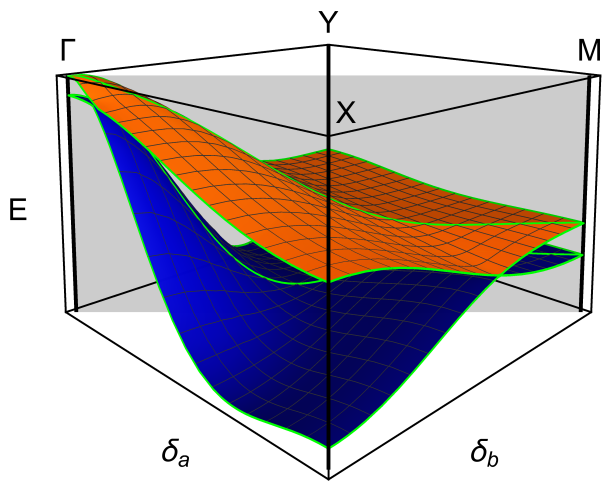}\\
      \caption{\label{fig:fig5}Two-dimensional representation of the two $p$-like TB bands in the restricted BZ of the primitive-rectangular lattice. A pair of \textit{conical intersections} near the $M$-point along the $M \rightarrow X$ line and near the $\Gamma$-point along the $\Gamma \rightarrow X$ line are now clearly visible. For this figure, interactions up to the eighth shell of Bravais-lattice neighbors have been included. Interaction parameters have been taken as $V^{(\I)} (pp\sigma)=1$, $V^{(\II)} (pp\sigma)=0.8$, $V^{(\III)} (pp\sigma)=0.7$, $V^{(\IV)} (pp\sigma)=0.5$, $V^{(\V)} (pp\sigma)=0.3$, $V^{(\VI)} (pp\sigma)=0.4$, $V^{(\VII)} (pp\sigma)=0.2$, $V^{(\VIII)} (pp\sigma)=0.1$, while $V(pp\pi)=-\frac{1}{10} V(pp\sigma)$ in all cases. The lattice constants have nominal values of $a=1.5$ and $b=1$.}
\end{center}
\end{figure}

This pair of conical intersections is clearly associated with band non-parabolicity, developing along low-symmetry edges of the restricted BZ, close to, but away from, $\Gamma$ and $M$ high-symmetry points. Furthermore, just as band non-parabolicity, the two conical intersections merge singularly into band-warping at $\Gamma$ and $M$ high-symmetry points when required to do so by the larger symmetry onset of the square Bravais lattice. Vice versa, for the primitive-rectangular Bravais lattice, the two conical intersections persist robustly for a wide range and choice of parameters. In fact, we have extended our $p$-type two-band TB model to include interactions up to the eighth shell of neighboring Bravais-lattice points, well beyond the third shell of neighbors to which \Figref{fig:fig3} and \Figref{fig:fig4} refer. With this extended choice of interactions and parameters, one conical intersection moves away from the $\Gamma \rightarrow Y$ line and shifts along the $\Gamma \rightarrow X$ line, while the other conical intersection drifts further away from the $M$-point along the $M \rightarrow X$ line. These results are displayed in \Figref{fig:fig5}.
Conical intersections are typically interpreted as Dirac points and they have become the object of major inquiry since the experimental realization in graphene and other kinds of topological insulators.\cite{CastroNoselovGeimRMP2009, BansilLinDasRMP2016} Two-dimensional graphite, or graphene, has a honeycomb structure, which is represented by an hexagonal Bravais lattice with two basis points in the primitive cell. Dirac points occur for $\pi$-bands at $K$ (or $P$) points, alternating inequivalently at corners of the hexagonal BZ. At a $K$-point, $\pi$-band degeneracy is still required by the lower symmetry of the $D_{3h}$ small-point-group of the corresponding $\bd{k}$ vector.\cite{Bassani} Original TB calculations in graphite and other layer compounds were performed in Refs.\ \onlinecite{BassaniPastoriNuovoCimento1967, DoniPastoriNuovoCimento1969}. 

The structure of Dirac points is induced and ``protected'' by symmetry in graphene and other topological insulators, whereas it is ``accidental'' in the conical intersections that we found in the primitive-rectangular Bravais lattice. This basic distinction may not be decisive, however. Topological considerations indicate that symmetry protection is not required to produce typical features of Dirac and Weyl points in two- and three-dimensional topological insulators and semimetals.\cite{HaldanePRL61-1988, ParkMarzariPRB84-2011, VanderbiltPRB92-2015} Our calculations further suggest that conical intersections of topological origin may be more common and robust than currently thought.

\section{\label{sec:NEB}Nearly-Free-Electron Approximation for Primitive-Rectangular and Square Lattices}
Although quite reasonable and consistent with experience, one may wonder whether the characteristics of electronic band warping that we have demonstrated might be partly due to peculiarities of TB approximations or underlying assumptions of orbital localization. It is thus appropriate to further consider an approximation that represents to some extent the opposite limit of wavefunction delocalization. That is of course the nearly-free-electron (NFE) approximation, which preferably applies to conduction bands higher in energy, especially for metals (see Ref.\ \onlinecite{Ashcroft}, Chp.\ 9, for example). 

Thus we begin with an ``empty lattice," traversed by a free-electron plane wave
\begin{equation}
\Psi^{(0)}_{\bd{k} + \bd{G}} (\bd{r}) = \frac{1}{\sqrt[]{V}} e^{i (\bd{k} + \bd{G}) \cdot \bd{r}}
\end{equation}
with perfectly parabolic energy
\begin{equation}
E^{(0)} (\bd{k} + \bd{G}) = \frac{\hbar^2}{2 m} \left( \bd{k} + \bd{G} \right)^2.
\end{equation}
Use of the reduced BZ scheme prepares the terrain for the introduction of a periodic crystal potential
\begin{equation}
V(\bd{r}) = \sum_{\bd{G}} U(\bd{G}) e^{i \bd{G} \cdot \bd{r}}
\end{equation}
in the empty lattice, requiring the Fourier-series coefficients to satisfy the relation 
\begin{equation}
U(\bd{G}) = U(-\bd{G})^{*}
\end{equation}
for $V(\bd{r})$ to be real. 
\begin{figure}[!hb]
	\begin{center}
 	\includegraphics[width=\figsize]{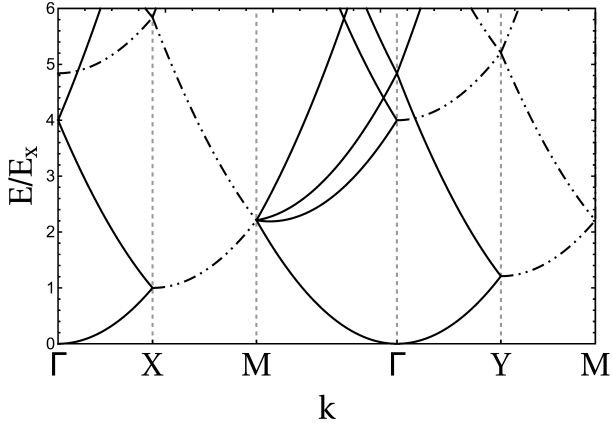}\\
    \caption{\label{fig:fig6} Empty lattice bands for a primitive-rectangular reciprocal lattice reduced to the first BZ.  Energies are expressed in units of the empty-lattice energy $E^{(0)}_1 \left(\tfrac{1}{2} \bd{g}_1 \right) = \tfrac{\hbar^2}{2 m } \left( \tfrac{\pi}{a} \right)^2 \equiv E_a$ at X. Insertions of double dots along a band denote double degeneracy throughout that band. Lattice constants are taken as $a= 1.1$ and $b=1$.}
\end{center}
\end{figure}
\begin{figure}[!hb]
	\begin{center}
    \includegraphics[width=\figsize]{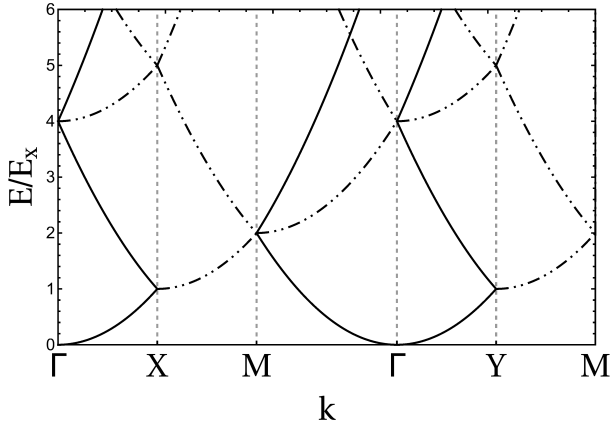}\\
    \caption{\label{fig:fig7} Empty lattice bands for a square reciprocal lattice reduced to the first BZ.  Energies are expressed in units of the empty-lattice energy $E^{(0)}_1 \left(\tfrac{1}{2} \bd{g}_1 \right) = E^{(0)}_1 \left(\tfrac{1}{2} \bd{g}_2 \right) = \tfrac{\hbar^2}{2 m } \left( \tfrac{\pi}{a} \right)^2 \equiv E_a$ at X or Y. Insertions of double dots along a band denote double degeneracy throughout that band. The lattice constant is taken as $a = b = 1$.}
\end{center}
\end{figure}

Folding consecutive BZ's of the extended BZ scheme into the reduced BZ scheme for a primitive-rectangular lattice, we obtain the band structure shown in \Figref{fig:fig6}. At the $\Gamma$-point ($\bd{k}=0$) we have the following sequence of one singly- and two doubly-degenerate empty-lattice energy levels 
\begin{subequations}
\begin{align}
E_1^{(0)} (\bd{0}) &= 0,\\
E_2^{(0)} (\bd{g}_1) &= E_2^{(0)} (-\bd{g}_1) = \frac{\hbar^2}{2 m } \left( \frac{2 \pi}{a} \right)^2,\\
E_3^{(0)} (\bd{g}_2) &= E_3^{(0)} (-\bd{g}_2) = \frac{\hbar^2}{2 m } \left( \frac{2 \pi}{b} \right)^2.
\end{align}
\end{subequations}
At the $X$-point ($\bd{k}= \tfrac{1}{2} \bd{g}_1 = \tfrac{\pi}{a} (1,0)$) we have a doubly-degenerate empty-lattice level with energy 
\begin{equation}\label{eqn:e0p}
E^{(0)}_1 \left(\tfrac{1}{2} \bd{g}_1 \right) = E^{(0)}_1 \left(-\tfrac{1}{2} \bd{g}_1 \right) = \tfrac{\hbar^2}{2 m } \left( \tfrac{\pi}{a} \right)^2 \equiv E_a .
\end{equation}
At the $M$-point ($\bd{k}=\tfrac{1}{2} \left( \bd{g}_1 + \bd{g}_2 \right) = \pi \left( \tfrac{1}{a}, \tfrac{1}{b} \right)$), we have a four-degenerate empty-lattice level with energy 
\begin{align}\label{eqn:e1p}
E_1^{(0)} \left( \frac{1}{2} \left( \pm \bd{g}_1 \pm \bd{g}_2 \right) \right) &= E_1^{(0)} \left( \frac{1}{2} \left( \pm \bd{g}_1 \mp \bd{g}_2 \right) \right) \nonumber\\
&=\frac{\hbar^2 \pi^2}{2 m} \left( \frac{1}{a^2} + \frac{1}{b^2} \right). 
\end{align}

Consideration of a slowly-varying periodic potential $V(\bd{r})$ enables us to consider only a few of its Fourier-series coefficients, namely,  
\begin{subequations}
\begin{align}
&U(\bd{g}_1) = U(-\bd{g}_1)^{*} \equiv U_1,\\
&U(\bd{g}_2) = U(-\bd{g}_2)^{*} \equiv U_2,\\
&U(2\bd{g}_1) = U(-2\bd{g}_1)^{*} \equiv U_{11},\\
&U(2\bd{g}_2) = U(-2\bd{g}_2)^{*} \equiv U_{22},\\
&U(\bd{g}_1 + \bd{g}_2) = U(-\bd{g}_1-\bd{g}_2)^{*} \equiv U_{12},\\
&U(\bd{g}_1 - \bd{g}_2) = U(-\bd{g}_1+\bd{g}_2)^{*} \equiv U_{12}.
\end{align}
\end{subequations}
The point-group of symmetry, $C_{2v}$, of the primitive-rectangular lattice further requires that these Fourier-series coefficients be real. Since the point-group $C_{2v}$ allows only one-dimensional irreducible representations, all degeneracies of the empty lattice should be removed by introduction of a potential $V(\bd{r})$ at all but a few exceptional points in reciprocal space. No band warping is expected at any resulting extremal non-degenerate point. 

Nevertheless, it is worth considering the band structure in the NFE approximation around the doubly-degenerate empty-lattice energies $E_2^{(0)}$ and $E_3^{(0)}$ at the $\Gamma$-point and the quadruply-degenerate empty-lattice energy $E_1^{(0)}$ at the $M$-point of a primitive-rectangular lattice.

Let us then apply a first-order perturbation theory to the energy doublet, originating from $E_2^{(0)}(\pm \bd{g}_1) \equiv 4 E_a$ at the $\Gamma$-point, at nearby points $\bd{k}_{\tilde{\Gamma}} = \pi \left(\tfrac{\delta_a}{a}, \tfrac{\delta_b}{b}\right)$, where $|\delta_a|, |\delta_b| \ll 1$, thus close to the $\Gamma$-point. This leads to diagonalization of a 2x2 matrix with elements $M_{11} = E^{(a)} (\bd{k}_{\tilde{\Gamma}}) - E^{(0)} (\bd{k}_{\tilde{\Gamma}} + \bd{g}_1)$, $M_{22} = E^{(a)} (\bd{k}_{\tilde{\Gamma}}) - E^{(0)} (\bd{k}_{\tilde{\Gamma}} - \bd{g}_1)$, and $M_{12} = M_{21} = -U(2 \bd{g}_1)$. Setting the matrix determinant equal to zero provides the eigenvalue difference
\begin{align}
&E_{+}^{(a)} (\bd{k}_{\tilde{\Gamma}}) - E_{-}^{(a)} (\bd{k}_{\tilde{\Gamma}}) =\nonumber\\ &\sqrt{ \left( E^{(0)} (\bd{k}_{\tilde{\Gamma}} + \bd{g}_1) -  E^{(0)} (\bd{k}_{\tilde{\Gamma}}-\bd{g}_1)  \right)^2 + 4 |U (2 \bd{g}_1)|^2}
\end{align}

The positive constant term, $4|U(2\bd{g}_1)|^2 > 0$, under the square root makes it always possible to perform a multi-dimensional Taylor series expansion, starting with a quadratic term, around the minimum energy difference or gap $E_g^{(a)} =  2|U(2\bd{g}_1)|$. That gap opens up at the $\Gamma$-point as a result of the slowly varying periodic potential perturbation. No band warping thus occurs for either one of the corresponding two non-degenerate bands, although they may exhibit greater and greater band non-parabolicity, the smaller and smaller becomes their minimum energy difference or gap $E_g^{(a)} = 2|U(2\bd{g}_1)|$.

An analogous result applies to the higher energy doublet originating from $E_3^{(0)}(\pm \bd{g}_2) = 4 E_a (\tfrac{a}{b})^2$ in the empty lattice at the $\Gamma$-point. Perturbation by a slowly varying periodic potential similarly opens a minimum energy gap $E_g^{(b)} = 2|U(2\bd{g}_2)|$ between two non-warped non-degenerate bands, quadratic most closely to that minimum energy gap, but with increasing band non-parabolicity further away, the more so the smaller is their minimum energy gap $E_g^{(b)} = 2|U(2\bd{g}_2)|$.

Let us now consider exactly a \textit{square lattice}. Corresponding empty lattice bands in the reduced BZ scheme are shown in \Figref{fig:fig7}. Based on the previous analysis with a TB model, we may expect a similar singular transition between the energy bands that we just obtained with a NFE model for a primitive-rectangular lattice near the $\Gamma$-point and the corresponding bands for a square lattice. Perhaps unexpectedly, however, this does not turn out to be so in this particular case. Within the same NFE approximation, let us then examine more closely the square lattice case, where $a = b$, $|\bd{g}_1| = |\bd{g}_2|$, and
\begin{subequations}
\begin{align}
E_2^{(0)} (\bd{g}_1) &= E_2^{(0)} (-\bd{g}_1) = \frac{\hbar^2}{2 m } \left( \frac{2 \pi}{a} \right)^2 = \\ E_2^{(0)} (\bd{g}_2) &= E_2^{(0)} (-\bd{g}_2) = \frac{\hbar^2}{2 m} \left( \frac{2 \pi}{a} \right)^2
\end{align}
\end{subequations}
%
form a quadruplet at the $\Gamma$-point in the empty lattice.

\begin{figure*}
	\begin{center}
    \includegraphics[width=12.9cm]{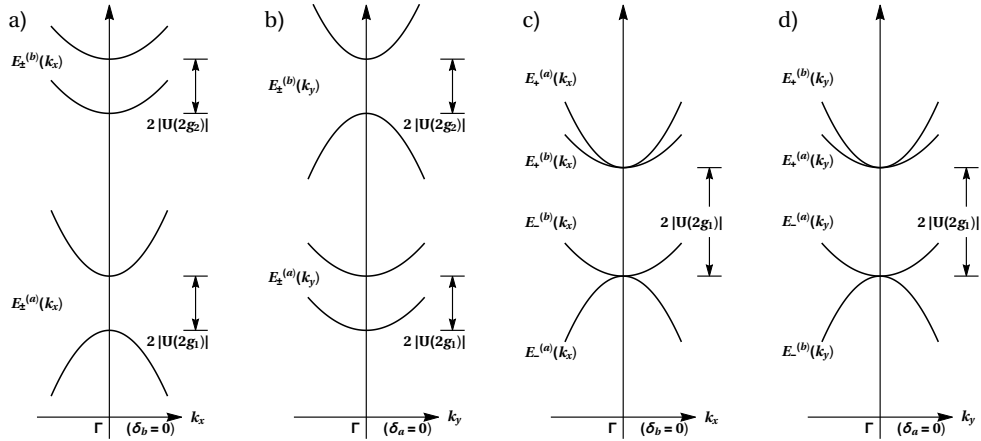}\\
    \caption{\label{fig:fig8}Schematic splitting of the lowest four (folded) bands in the empty lattice near the $\Gamma$-point by introduction of a slowly varying periodic potential in the primitive-rectangular lattice, a) and b), and in the square lattice, c) and d). In this case, due to the simplicity of the assumed periodic potential, there is simply a continuous non-warping transition between the four non-degenerate bands of the primitive-rectangular lattice and the two doubly-degenerate bands at $\Gamma$ of the square lattice. Namely, there is no removal of smoothness and consequent band warping produced by the greater symmetry and degeneracy required for the square lattice.}
\end{center}
\end{figure*}

Perturbation by a slowly varying periodic potential partly splits this quadruplet into a pair of still degenerate doublets at the $\Gamma$-point, now separated by a single energy gap $E_g = 2|U(2\bd{g}_1)| = 2|U(2\bd{g}_2)|$. At $\bd{k}_{\tilde{\Gamma}} = \tfrac{\pi}{a} \left( \delta_a, \delta_b \right)$ points close to the $\Gamma$-point one can still solve analytically a 4x4 determinant equation for the perturbed quadruplet, because the 4x4 matrix factorizes diagonally into two 2x2 block-matrices, when only $U(2\bd{g}_1) = U(2\bd{g}_2)$ interactions are considered for the square lattice in the NFE approximation. Both eigenvalues of each of the two 2x2 block-matrices involve square roots containing within them the same positive constant, $E_g^2 = 4|U(\bd{g}_1)|^2 > 0$. That always allows a Taylor series expansion, starting with a quadratic term, hence, no band warping occurs. Thus, somewhat unexpectedly, we have in this NFE approximation for a square lattice two pairs of bands that become degenerate smoothly, i.e., without generation of any band warping at the $\Gamma$-point, where each doublet is separated from the other by a single energy gap $E_g = 2|U(2\bd{g}_1)|$. Each doublet's degeneracy is required by the greater symmetry point-group $C_{4 v}$ at the $\Gamma$-point. This smooth touching and degeneracy of each pair of bands at the $\Gamma$-point can thus be continuously derived from the corresponding quadruplet of non-degenerate bands of a primitive-rectangular lattice by letting $a$ approach $b$ continuously. For a primitive-rectangular lattice, no degeneracy of the quadruplet at the $\Gamma$-point and the presence of two separate and different energy gaps, $E_g^{(a)} = 2|U(2\bd{g}_1)|$ and $E_g^{(b)} = 2|U(2\bd{g}_2)|$, is consistent with the reduced symmetry of $C_{2 v}$, which admits no irreducible representation of dimension greater than one. This continuous non-warping transition of the quadruplet bands along the $k_x$ and $k_y$ axes around the $\Gamma$-point from the primitive-rectangular to the square lattice is depicted in \Figref{fig:fig8} a) to d). 

By contrast, after considering more complex models and crystal structures, we much more frequently found that bands becoming degenerate at extremal energy points in the BZ become typically warped as a result of symmetry requirements. Thus, the elementary example that we have just illustrated, displaying \textit{non-warped degenerate bands} at the $\Gamma$-point, is rather uncommon and essentially due to the simplicity of the NFE potential approximation that we have considered for the square lattice in this instance. Nevertheless, such possibility cannot be excluded in general, and it is not nearly as unusual as its converse, namely, that of a \textit{non-degenerate band becoming warped} at an energy extremum as a result of more deliberate artifacts. So, let us proceed to demonstrate that perturbation with a more elaborate periodic potential diagonalization in a NFE model can and does produce a singular symmetry transition to a pair of degenerate and warped bands. That is the case for the lowest quadruplet of bands at the corner of the BZ, i.e., at the $M$-point, in passing from the primitive-rectangular to the square lattice. 

Namely, let us consider a point $\bd{k}_{\tilde{M}} = \pi \left( \tfrac{1-\delta_a}{a}, \tfrac{1-\delta_b}{b} \right)$ in a neighborhood of the $M$-point, thus having $|\delta_a|,|\delta_b| \le 1$. Now we have to consider a full 4x4 matrix representing the interaction of four plane waves with wave vectors $\bd{k}_{\tilde{M}}$,  $\bd{k}_{\tilde{M}} - \bd{g}_1$,  $\bd{k}_{\tilde{M}} - \bd{g}_2$, and $\bd{k}_{\tilde{M}} - \bd{g}_1 - \bd{g}_2$. To demonstrate the essential origin of band warping, it is enough to consider a slowly varying periodic potential with three Fourier components, namely $U(\bd{g}_1) \equiv U_1 >0, U(\bd{g}_2) \equiv U_2 > 0$, and $U(\bd{g}_1 -\bd{g}_2)=U(\bd{g}_2-\bd{g}_1)=U(\bd{g}_1 + \bd{g}_2) = U(-\bd{g}_1-\bd{g}_2) \equiv U_{12} > 0$. These three parameters, $U_1$, $U_2$, and $U_{12}$, all differ from one another in a primitive-rectangular lattice. That removes all degeneracies of the quadruplet represented in \Eq{eqn:e1p} for the empty lattice at the $M$-point, adding the energies $U_1-U_{12}-U_2$, $-U_1+U_{12}-U_2$, $-U_1 - U_{12} +U_2$, and $U_1  + U_{12} + U_2$ to each level. Accordingly, we obtain four non-degenerate non-warped bands at the $M$-point of a primitive-rectangular lattice, consistent with its $C_{2v}$ symmetry. However, for a square lattice with $a = b$, $U_{1} = U_{2}$ is required by the greater symmetry of the $C_{4v}$ point-group, doubled in size and thus admitting two-dimensional irreducible representations. As a result, the two middle bands become degenerate at the $M$-point of a square lattice and they become correspondingly warped therein. The origin of this warping is due to the canceling of zero-order or constant terms under two nested square roots. It is thus worth describing in some detail this more elaborate mechanism of cancellation.

\begin{figure}[!hb]
	\begin{center}
 	\includegraphics[width=\figsize]{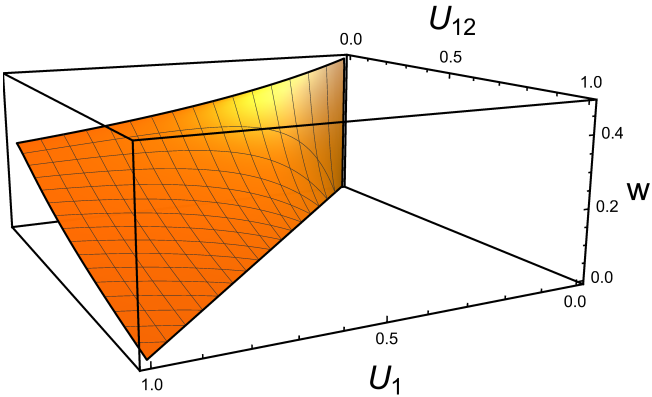}\\
      \caption{\label{fig:fig9} Warping parameter, $w$, for the upper middle band, $E_+$, having a relative minimum at $M$ for the square lattice.  $U_1$ and $U_{12}$ are given in units of $E_a$, and $a = 1$ has been set.}
\end{center}
\end{figure}

The angular effective mass procedure to evaluate warping of the two middle degenerate bands is to expand each one in a one-dimensional \textit{radial} Taylor series, which is generally possible.\cite{MRPF} These bands thus become represented as 
\begin{equation}\label{eqn:Epm}
E_{\pm}(k_r, \theta) = E_0 + f_{2 \pm} (\theta) k_r^2 + \ldots
\end{equation}
where one can show that
\begin{equation}
f_{2 \pm} (\theta) = a \pm \sqrt{b (1+\cos 4 \theta) + c (1- \cos 4 \theta)}.
\end{equation}
The $a$, $b$, and $c$ parameters are given by 
\begin{align}
a &= E_a + \frac{2 E_a^2 U_{12}}{U_1^2 - U_{12}^2},\nonumber\\
b &= \frac{2 E_a^4}{(U_1^2 - U_{12}^2)^2} U_1^2, \\
c &= \frac{2 E_a^4}{(U_1^2 - U_{12}^2)^2} U_{12}^2,
\end{align}
where $E_a = \tfrac{\hbar^2 \pi^2}{2 m a^2}$ is again defined as the energy unit. The warping parameter,\cite{MRPF} $w$, can be derived analytically as
\begin{widetext}
\begin{equation}
w_{\pm}(a,b,c) = \frac{8 \sqrt{(b+c) \pi^2 - 2 \left( c \,  \textrm{K}(1-\frac{b}{c})^2 + 2 \sqrt{b c} \, \textrm{K}(1-\frac{b}{c}) \, \textrm{K}(1-\frac{c}{b}) + b \, \textrm{K}(1-\frac{c}{b})^2 \right) }}{8 a \pi \pm 8 \sqrt{2} \left( \sqrt{c} \, \textrm{K}(1-\frac{b}{c}) + \sqrt{b} \, \textrm{K}(1-\frac{c}{b}) \right)},
\end{equation}
\end{widetext}
where $\textrm{K}(m)$ is the complete elliptic integral of the first kind, namely,
\begin{equation}
\textrm{K}(m) = \int_0^{\pi/2} \frac{\dd \theta}{\sqrt{1 - m \sin^2 \theta}}.
\end{equation}
A plot of this warping parameter, $w$, at $M$, as a function of the perturbing potential parameters $U_1$ and $U_{12}$, is shown in \Figref{fig:fig9} for the upper band in \Eqn{eqn:Epm}, $E_+$, which has a relative minimum at $M$. We made the reasonable assumption that $U_1 > U_{12} >0$ for a slowly varying potential, but an analogous plot can be obtained for $U_{12} > U_1 >0$. In either case, our plots indicate that warping of the two degenerate middle bands at $M$ are robust as $U_1$ and $U_{12}$ vary.  
\begin{figure}[!htp]
	\begin{center}
 	\includegraphics[width=\figsize]{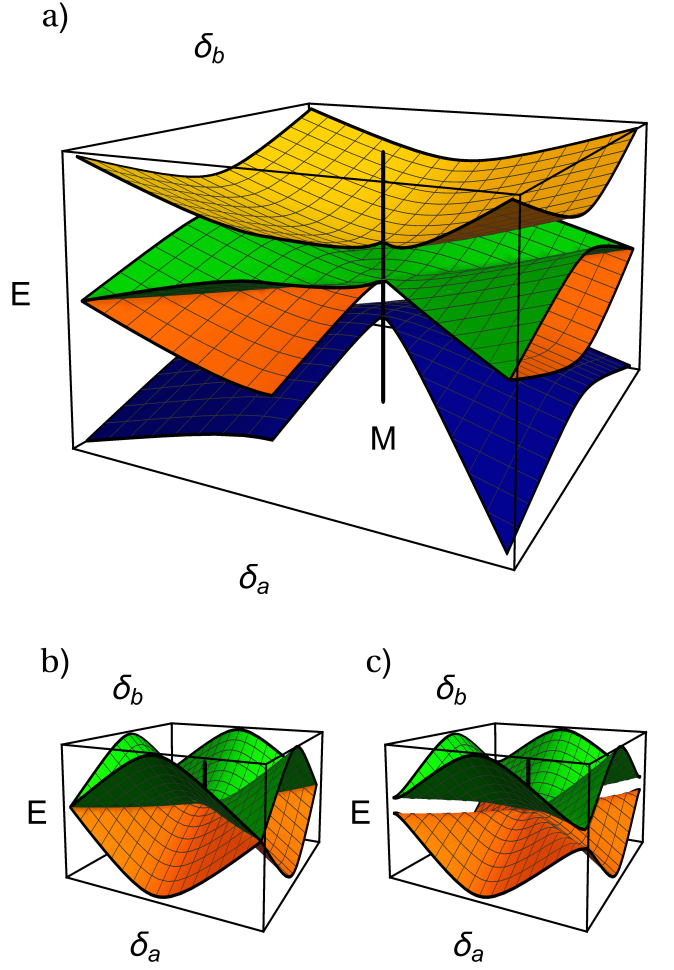}\\
      \caption{\label{fig:fig10} Lowest band structure near the $M$-point in the NFE model for the square lattice. Part (a) shows the splitting, mixing and warping of the whole quadruplet of bands, for $U_{12} = 0$. Part (b) focuses on the mixing and warping of the two middle bands degenerate at $M$, still for $U_{12} = 0$. Part (c) shows the splitting of the two middle bands around $M$ for $U_{12} \rightarrow 0$, but not identically zero: the accidental degeneracy is thus removed along the two diagonal lines off $M$.}
\end{center}
\end{figure}
For further demonstration, let us study the limiting case of $U_{12} \rightarrow 0$, while keeping finite $U_1 = U_2 >0$. This illustrates a similar, but more involved mechanism of band warping than that previously demonstrated in \Eqn{eqn:TB3} for the simpler case of just two $p$-like bands treated with a TB model. Now, for the upper middle band, $E_+$, degenerate at $M$ with the lower middle band, $E_-$, in \Eqn{eqn:Epm}, the energy dispersion can be transformed back in terms of rectangular coordinates as 
%
%
%
%
%
%
\begin{align}\label{eqn:quarticradicalization}
&E_+(\delta_a, \delta_b) = E_a \left(\delta _a^2+\delta _b^2+2\right)+\sqrt{2} \times \nonumber\\&
   \sqrt{2 E_a^2 \left(\delta _a^2+\delta _b^2\right)+
   U_1^2 - \sqrt{\left(4 \delta_a^2
   E_a^2+U_1^2\right) \left(4 \delta_b^2 E_a^2+U_1^2\right)}}.
\end{align}
%
%
%
%
%
%
In this expression, \Eqn{eqn:quarticradicalization}, if we let $\delta_a$ and $\delta_b$ vanish, there remains a positive constant, $U_1^4$, under the inner square root. That would permit a Taylor series expansion of the inner square root all by itself. However, subtraction of the square root of that constant from another constant, $U_1^2$, which belongs to the outer square root, eliminates any possibility of a Taylor series expansion for the overall quartically radicalized expression and mixing of the two middle bands degenerate at $M$. In \Figref{fig:fig10} (a), we show the splitting, mixing and warping for the quadruplet of bands at $M$.  In \Figref{fig:fig10} (b), we focus on the mixing and warping of the two middle bands degenerate at $M$. The line-degeneracy appearance must be further investigated, however. Having set $U_{12} = 0$ identically is causing the two middle bands to remain accidentally degenerate along two diagonal lines intersecting at $M$. That would permit a labeling of those bands as two saddle-point bands at $M$, rotated by $90^{\circ}$ degrees relative to each other, thus crossing each other along those two diagonal stationary lines. However, such labeling, which derives from the original primitive-rectangular lattice, becomes ultimately incorrect for the square lattice. The reason is that, in the limit of a vanishing $U_{12} \rightarrow 0$, but not identically zero, the accidental degeneracy is removed along the two diagonal lines off $M$ in the square lattice, as generally expected.\cite{landauLifshitzQM} Degeneracy can and must be maintained exclusively at $M$, as an isolated singular point, because that is required by the point-group symmetry, $C_{4v}$, for the square lattice. In \Figref{fig:fig10} (c) we show the splitting of the two middle bands everywhere around $M$ for a small, but non-vanishing $U_{12}$. The corresponding analytic expression is too cumbersome to be reported here, but it is not essentially different from that reported in \Eqn{eqn:quarticradicalization}, where we have carried out the proper limit of $U_{12} \rightarrow 0$. Therefore, only \Eqn{eqn:quarticradicalization} properly corresponds to the correct limiting procedure and band labeling appropriate to the square lattice, insofar as the analysis of second-order non-differentiability and band warping is concerned. This is also confirmed by examining the posterior rectangular face shown in \Figref{fig:fig9}, which correspondingly indicates a non-zero warping parameter at the $M$-point of the square reciprocal lattice in the limit of $U_{12} \rightarrow 0$, associated with the upper band in \Eqn{eqn:Epm} and \Eqn{eqn:quarticradicalization}, $E_+$, which can only have a relative minimum at the singular point of contact with $E_-$, at $M$. 

Other complex mechanisms of band warping are possible or can be envisioned. However, the basic examples that we have analyzed in this paper provide a central understanding and insight about various possibilities and mechanisms of warping, particularly for degenerate bands at symmetry points in reciprocal space, from a fundamental perspective.

\section{\label{sec:Conclusion}Conclusion}
We have demonstrated from a fundamental perspective the physical and mathematical origin of ``band warping" in electronic band structures using two fundamental and complementary approaches to their calculation, namely, the tight-binding (TB) method and the nearly-free-electron (NFE) approximation. We have concluded that for the majority of \textit{non-degenerate} bands with energy extrema one should generally expect no band warping, irrespective of whether band non-parabolicity may or may not be prominent. Band warping is typically and ultimately a consequence of the singular onset of a larger symmetry group, allowing the possibility of irreducible representations of higher dimensions at special symmetry points in reciprocal space.

For the sake of illustration of basic principles, we have mainly discussed in this paper results for two fundamental crystal structures, namely, those of two-dimensional primitive-rectangular and square Bravais lattices, which can be investigated analytically to a remarkable extent. In fact, we have further investigated many more complex structures and we have typically found that band-warping and band non-parabolicity features maintain the basic relations and characteristics demonstrated in this paper.

From a mathematical perspective, perhaps the most crucial feature for the appearance of band warping is the cancellation of positive constants under square roots in the process of matrix diagonalization for interacting levels that must become degenerate at isolated critical points in reciprocal space as required by crystal symmetry. That cancellation of constants under square roots eliminates the possibility of a multi-dimensional Taylor series expansion at a critical point beyond its zero first-order differential. We have demonstrated this process at the quadratic level of radicalization with a TB model for a square lattice, and at a quartic level of radicalization within a NFE approximation for the same square lattice. 

This analysis, based on multi-dimensional Taylor series expansions in rectangular coordinates, is complemented, and in fact resolved, by resorting to one-dimensional \textit{radial} Taylor series expansions, through a method that involves the determination of a correspondingly more general and applicable concept of angular effective mass. That allows computation of a band warping parameter, $w$, which provides some measure of how severely an energy surface is forbidden from having smooth curvature or second-order differentiability at an isolated critical point. We have computed non-zero $w$'s to confirm warping of bands becoming degenerate at critical points as required by the square lattice symmetry.

In the process of examining the effects of band non-parabolicity, we have discovered a phenomenon of conical intersections of two $p$-type bands that are required to go underneath and reemerge from one another by the global topology of a primitive-rectangular Bravais lattice, studied with a fundamental TB model. Its conical intersections have ultimately the same structure as Dirac points, extensively studied in graphene and other kinds of topological insulators. Their origin is quite different, however. The degeneracy of conical intersections at ``strangulated'' singular points is ``accidental'' in our primitive-rectangular Bravais lattice, whereas it is required and ``protected'' by symmetry at Dirac points of more complex materials and lattice structures. Nevertheless, our conical intersections are quite robust under variation of TB parameters and inclusion of interactions among several shells of Bravais-lattice neighbors. Conical intersections of topological origin may thus be more common and less peculiar in two- and three-dimensional structures than currently assumed for Dirac and Weyl points in topological insulators and semimetals. 

\acknowledgments
This work was supported by the Vitreous State Laboratory of The Catholic University of America.

%

\end{document}